\documentclass[%
superscriptaddress,
twocolumn,
amsmath,amssymb,
aps,
prl,
floatfix
]{revtex4}
\usepackage{graphicx}
\usepackage{graphics}
\usepackage{dcolumn}%
\usepackage{epsfig}
\usepackage{bm}
 
\usepackage{color}

\newcommand {\tphi}{\tilde{\phi}}
\newcommand {\tPhi}{\tilde{\Phi}}

\newcommand {\tchi}{\tilde{\chi}}

\newcommand{\p}{{\cal P}}
\newcommand{\PT}{{\cal PT}}
\newcommand{\T}{{\cal T}}

\newcommand{\tvarphi}{\tilde{\varphi}}

\newcommand{\bPsi}{\bm{\Psi}}

\newcommand{\bzeta}{ \mbox{\boldmath$\zeta$\unboldmath}}

\newcommand{\bv}{{\bm v}}

\newcommand{\rA}{{\rm A}}

\newcommand{\zpb}{{z_{\rm pb}}}

\newcommand{\rev}[1]{\textcolor{black}{#1}}
\newcommand{\rerev}[1]{\textcolor{black}{#1}}

\begin{document}

\title{
%Nonlinear Schr\"odinger Equations with Non-Hermitian Gauge Potentials: 
\rev{Superexponential Amplification, Power Blowup, and Solitons sustained by Non-Hermitian Gauge Potentials}}

\author{Dmitry A. Zezyulin}
\affiliation{ITMO University, St. Petersburg 197101, Russia}

\author{Yaroslav V. Kartashov}
\affiliation{Institute of Spectroscopy, Russian Academy of Sciences, Troitsk, Moscow, 108840, Russia}

\author{Vladimir V. Konotop}
\affiliation{Departamento de F\'{i}sica and Centro de F\'{i}sica Te\'orica e Computacional, Faculdade de Ci\^encias, Universidade de Lisboa, Campo Grande, Ed. C8, Lisboa 1749-016, Portugal}
\email{vvkonotop@fc.ul.pt}

\date{\today}

\begin{abstract}
 \rerev{We introduce a continuous one-dimensional non-Hermitian matrix gauge potential and study its effect on dynamics of a two-component field. The model is emulated by} a system of evanescently coupled nonlinear waveguides with distributed gain and losses. The considered gauge fields lead to a variety of unusual physical phenomena in both linear and nonlinear regimes. In the linear regime, the field may undergo superexponential convective amplification. A total power of an input Gaussian beam may exhibit a finite-distance blowup, which manifests itself in absolute delocalization of the beam at a finite propagation distance, where the amplitude of the field remains finite. The defocusing Kerr nonlinearity initially enhances superexponential amplification, while at larger distances it suppresses the growth of the total power. The focusing nonlinearity at small distances slows down the power growth and eventually leads to the development of the modulational instability. Complex periodic gauge fields  lead to the formation of families of stable fundamental and dipole solitons.

\end{abstract}
\maketitle

Impact of imaginary magnetic fields or, more generally, of non-Hermitian gauge potentials on wave processes attracts considerable attention nowadays. This interest was initially motivated by the discovery of localization transitions and mobility edges in random systems due to imaginary vector potentials~\cite{Hatano1996,Hatano1997}. The unusual features introduced by such potentials into physics of wave localization were further investigated in~\cite{Brouwer,Goldsheid,Takeda,Heinrichs,CompGauge2D}. More recently, it was shown that non-Hermitian gauge potentials can support robust transport in chains with non-Hermitian hopping~\cite{LonghiGattiPRB,LonghiGattiSciRep,Longhi2017PRA}, and enhance forces acting on photons~\cite{Lana}. Complex vector potentials were also introduced for non-Hermitian extensions of the Dirac equation~\cite{Takeda,Alexandre,Nori}, where they result in Lorentz-symmetry violation.

In contrast to Hermitian gauge fields that are present in description of different physical systems and  can even be designed at will, for example, in atomic systems~\cite{Ruseckas,GaugeReview}, applications of non-Hermitian gauge fields and approaches to their creation remain scarce. In Refs.~\cite{Hatano1996,Hatano1997} the imaginary magnetic field was introduced in the context of imaginary-time description of localization of bosons in superconducting vortex arrays \rev{\cite{NelVin,Nelson1990}}. Models emulating non-Hermitian gauge fields by complex hopping between neighboring sites were proposed using optical settings, such as coupled microring resonators~\cite{LonghiGattiPRB}, photonic lattices~\cite{LonghiGattiSciRep}, and frequency lattices~\cite{Quin}. Non-Hermitian arrays emerging in such models are linear, discrete, and characterized by a scalar field. Implementation of the effective imaginary gauge field in a system of parallel slabs using non-reciprocal elementary cells consisting of microrings or nanoparticles was also suggested~\cite{Lana}.

\rerev{In this Letter, we introduce a continuous one-dimensional (1D) {\em non-Hermitian matrix} gauge potential and study its effect on linear and nonlinear {\em spinor} fields.}  We show that such potentials enable new striking features of the dynamics, ranging from unconventional superexponential convective amplification, to \rev{(linear)} power blowup leading to complete delocalization at a finite propagation distance, and formation of stable soliton complexes in periodic non-Hermitian gauge fields.

\begin{figure}%[t]
\begin{center}
\includegraphics[width=0.9\columnwidth]{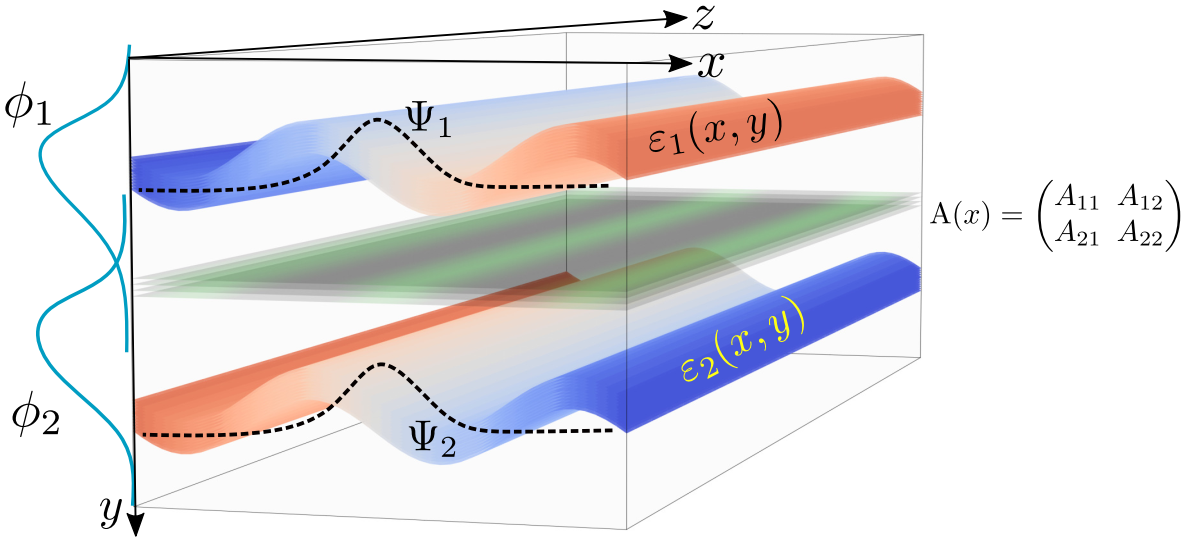}
\end{center}
\caption{Schematics of two waveguides (shown by the wavy structures) with complex-valued dielectric permittivities $\varepsilon_{1,2}(x,y)$, which are separated in the $y$-direction. Each waveguide supports its own mode, $\phi_1(x,y)$ and $\phi_2(x,y)$, localized around minima of the effective potential wells, whose coupling creates an $x$-dependent complex-valued matrix gauge potential $\rA(x)$ for the envelopes $\Psi_{1,2}(x,z)$  propagating along the $z$-axis. }
	\label{fig:idea}
\end{figure}

\paragraph{Optical potential.} \rerev{We start by presenting  
 a simple optical system that allows to emulate the matrix gauge potential in experimentally feasible conditions. To this end} we consider a paraxial light beam propagating along the $z$-direction in a system of \rerev{two} evanescently coupled waveguides which are separated along the $y$-axis [as illustrated schematically in Fig.~\ref{fig:idea}]. \rerev{The waveguides have gain and losses  characterized by a differentiable function $\eta(x)$ which is bounded, $|\eta(x)|\leq h_0$, by a constant $h_0$. The transverse dielectric permittivity of such a structure is described by a nonseparable optical potential $V(x,y)$. We assume it to be of the form $V(x,y)\equiv V_0(y+i\eta(x))$, where $V_0(y)$ is a real even double-well potential describing the waveguides without gain and losses: $\p_yV_0(y):=V_0(-y)=V_0(y)$. Thus, $V(x,y)$ is $\p_y\T$-symmetric: $\p_y\T V(x,y) := V^*(x, -y)= V(x,y)$, where asterisk means complex conjugation. For most of the phenomena considered below $\eta$ is considered small enough and thus $V\approx V_0(y)+i\eta(x)V_0'(y)$. While a double-well potential is the most standard model for dual-core optical waveguides, required distributions of gain and losses can be created by doping of its cores with active impurities, by shaping pump beam in nonlinear process providing gain~\cite{Kip}, or created in atomic cells filled by gasses of multilevel atoms that allow designing dielectric permittivity landscapes practically at will~\cite{Chao,MinXiao,ChaoKon,YiqiXiao}.}

\rerev{
To further specify the requirements on the potential, we consider an eigenvalue problem $H_\eta\tphi_{j}(x,y)=-\beta_{j}\tphi_{j}(x,y)$, where $H_\eta:= -({1}/{2})\partial_y^2+V(x,y)$ and $\tphi_{j}(x,y)$ are the eigenmodes localized along the $y$-axis. It will be assumed that the two eigenvalues with largest absolute values,
$\beta_{1}>\beta_2$, are {\em real} and $x$-independent (the last condition is for simplicity and can be relaxed) while superpositions   $\phi_j=({\tphi_2+(-1)^j\tphi_1})/{\sqrt{2}}$, $j=1,2$,  are localized in the cores of different waveguides with exponentially small overlap.
Assuming that there are no exceptional points in the spectrum of $H_\eta$, we  introduce the eigenmodes of the Hermitian adjoin: $H_\eta^\dag \tvarphi_{1,2}=-\beta_{1,2} \tvarphi_{1,2}$, such that $\langle \tvarphi_k,\tphi_j\rangle:=\int_{-\infty}^\infty \tvarphi_k^* \tphi_j dy=\delta_{kj}$, as well as their superpositions
$\varphi_j=({\tvarphi_2+(-1)^j\tvarphi_1})/{\sqrt{2}}$. 
}

\rerev{ Let $\tPhi_{1,2}(y)$ be real orthonormal eigenmodes of the Hermitian (i.e., $\eta(x)\equiv 0$) eigenvalue problem $H_0\tPhi_{1,2}=-\beta_{1,2} \tPhi_{1,2}$. Then $\p_y \tPhi_1(y)=\tPhi_1(y)$ and $\p_y \tPhi_2(y)=-\tPhi_2(-y)$. Let also 
$\tPhi_{1,2}(y)$ have analytic continuations from the real axis $y\in\mathbb{R}$ to a stripe Im$\,y \in (-h_0, h_0)$, and for any $h\in (-h_0, h_0)$ the eigenmodes remain localized in $y$: $\lim_{y\to \pm \infty} \tPhi_{1,2}(y+ih)=0 $. The same is also assumed for the derivatives $\tPhi_{1,2}'(y)$.  Then $\tphi_{1,2}(x,y) =\tPhi_{1,2}\left(y+i\eta(x)\right)$ are the eigenmodes of $H_\eta$ with the real and $x$-independent eigenvalues $\beta_{1,2}$. Moreover, all integrals   $\langle \tvarphi_{k}, \tphi_{j}\rangle$ do not depend on $\eta$ (and hence on  $x$) \cite{supplemental},  and $\phi_{1,2}$ satisfy the  biorthonormality conditions  for all $x$:   $\langle \varphi_k,\phi_j\rangle=\delta_{kj}$.
}

\rerev{
An example of the double-well potential (that we use below in numerics) is given by~\cite{razavy1980}
$V_{\rm ex}(y) =  \xi^2 f^2\cosh(4fy) - 4\xi f^2\cosh(2fy)$. Its eigenfunctions are known in the explicit form~\cite{supplemental}. The imaginary part of corresponding complex potential $V_\textrm{ex}(y+i\eta(x))$ produced at $\eta(x) \ll 1$ is perfectly compatible with experimentally achievable optical gain levels of a few $\textrm{cm}^{-1}$ \cite{Kip,YiqiXiao}.
}

\paragraph{Non-Hermitian gauge potential.} Further we consider the propagation of the paraxial beam in the medium in the presence of the above potential $V$ and Kerr nonlinearity, described by the nonlinear Schr\"odinger equation
\begin{equation}
\label{opt-eq1}
i\frac{\partial \Psi}{\partial z}= -\frac{1}{2}\nabla^2 \Psi+V(x,y)\Psi
+{U(x,y)}\Psi 
+\chi\rev{(x)} |\Psi|^2\Psi,
\end{equation}
where $\Psi$ is the dimensionless field amplitude, $\nabla\equiv (\partial_x,\partial_y)$, and $\chi\rev{(x)}$ is a  real function describing (generally speaking, $x-$dependent) Kerr coefficient. In (\ref{opt-eq1}) we introduced  an auxiliary potential $U(x,y)$ (whose role is specified below).

Now we employ the  two-mode approximation and look for a solution of (\ref{opt-eq1}) in the form $\Psi\approx e^{i(\beta_1+\beta_2)z/2 }[\Psi_1(x,z) \phi_1 +\Psi_2(x,z) \phi_2]$, where $\Psi_{1,2}$ are slowly-varying envelope amplitudes. Using this ansatz in Eq.~(\ref{opt-eq1}), applying   $\langle\varphi_j,\cdot\rangle$, and neglecting all nonlinear terms with integrals containing products of $\phi_{1}$ and $\phi_2$ (which is justified by their localization) we arrive at the equation for the column-vector $\bPsi=(\Psi_1,\Psi_2)^{\rm T}$:
 \begin{equation}
 	\label{GPE}
 	i\frac{\partial \bPsi}{\partial z}=\frac{1}{2}\Pi^2  \bPsi-\mathfrak{U} \bPsi + \left(\begin{array}{cc}
 		 \tchi_1|\Psi_1|^2 & 0 \\ 0 & \tchi_2|\Psi_2|^2
 	\end{array}\right)\bPsi.
 \end{equation} 
Here $\Pi=-i \partial_x -\rA(x)$, $\rA(x)$ is a complex-valued $2\times 2$ matrix gauge-potential with the entries $A_{kj}=\langle\varphi_{k},i\partial_x\phi_j\rangle$, $\tchi_j=\langle \varphi_j, \chi |\phi_j|^2\phi_j\rangle$ are the effective nonlinearity coefficients, $\mathfrak{U}=( \beta_2- \beta_1)\sigma_1/2 +{\rm u}(x)+\rA^2(x)/2$ is the effective matrix potential, ${\rm u}(x)$ is a $2\times 2$ matrix with entries $u_{kj}=\langle\partial_x\varphi_{k},\partial_x\phi_j\rangle/2\rev{-\langle\varphi_{k},U\phi_j\rangle}$, and $\sigma_{1,2,3}$ are the Pauli matrices.  For the $\PT$-symmetric double-well potential specified above  the gauge potential is obtained explicitly
 \begin{equation}
 	\label{A-minimal}
 	\rA(x)=i\eta_x \alpha\sigma_2, \qquad \alpha=\int_{-\infty}^{\infty}\tPhi_1'(y)\tPhi_{2}(y)dy. 
 \end{equation}
Here $\eta_x = d\eta(x)/dx$.

\rerev{Since the main goal of this paper is to describe  the effects that emerge specifically due to the non-Hermitian gauge, we observe that  the $\mathfrak{U}$ in (\ref{GPE}) can be made exactly zero by a judicious choice of the auxiliary potential $U(x,y)$~\cite{supplemental}. Alternatively,} one can consider smooth functions $\eta(x)$ allowing one to keep the terms $\sim\eta_x$ and neglecting those $\sim\eta_x^2$. Since for sufficiently large separation between the waveguides one can also achieve $ \beta_1- \beta_2\lesssim \eta_x^2$, the mismatch   between the propagation constants can be neglected, as well. Then, all entries of the matrix $\mathfrak{U}$ in Eq.~(\ref{GPE}) become of the order of $\eta_x^2$ and one can neglect $\mathfrak{U}$ even at $U(x,y)=0$. Therefore, from now on we consider the cases where $\mathfrak{U}=0$.

 \paragraph{Superexponential amplification.} In the linear limit,  $\tchi_1=\tchi_2=0$, the matrix gauge potential can be gauged out.  To this end we introduce time-independent mutually orthogonal carrier states $\bzeta_{1,2}(x)$, $\bzeta_1^\dagger \bzeta_2=0$, as solutions of the equation  $\Pi\bzeta_{1,2}(x)=0$ \cite{KKMS,KarKon20},  and  look for the field in the form $\bPsi(x,z)=v_1(x,z)\bzeta_1(x)+v_2(x,z)\bzeta_2(x)$, where $v_{1,2}(x,z)$ are the envelopes. For the gauge field (\ref{A-minimal}) we have 
\begin{eqnarray}
 \label{carrier}
 \bzeta_{1}=e^{\alpha\eta(x)}(-i,1)^{\rm T}, \quad  \bzeta_{2}=e^{-\alpha\eta(x)}(i,1)^{\rm T},
\end{eqnarray}
and the linear model reduces to  $i\bv_z=-(1/2)  \bv_{xx}$, 
where $\bv=(v_1,v_2)^{\rm T}$. Therefore, linear propagation  of the initial field distribution $\bPsi_0(x)$ can be solved explicitly:
 \begin{eqnarray}
 	\label{eq:linear}
 	\bPsi
 	=\frac{e^{-i\pi/4}}{\sqrt{2\pi z}}  \int_{-\infty}^{\infty}\!\! e^{i(x-\xi)^2/(2z)} 
 	\left(
 	\cosh\left\{\alpha \left[\eta(x)-\eta(\xi)\right]\right\}
 	\right. \nonumber \\ \left.	
 	+\sigma_2
 	\sinh\left\{\alpha \left[\eta(x)-\eta(\xi)\right]\right\}
 	\right)
 	\bPsi_0(\xi)d\xi. %\quad
 \end{eqnarray}
Thus, the gauge field directly affects the intensity distribution of the diffracting beam by rotating the input field $\Psi_0$ through an imaginary angle $i\alpha[\eta(x)-\eta(\xi)]$ in the transverse plane.
This may lead to unusual propagation scenarios. We describe them for  an input carrier state $\bzeta_1$ with a Gaussian envelope, $\Psi_0=e^{-x^2}\bzeta_1$, of the width $1/\sqrt{2}$.
Then (\ref{eq:linear})  becomes
\begin{equation}
	\label{init}
	\bPsi=\frac{-i}{\sqrt{1+ 2 i z}}\exp\left[\alpha \eta(x)-\frac{2 x^2}{2(1+i 2 z)}\right]\left(\begin{array}{cc}
		1 \\ i
	\end{array}\right).
\end{equation} 

Starting with an example of a $2\pi$-periodic function $\eta(x)=\eta(x+2\pi)$ and using the expansion $e^{2\alpha\eta}=  \sum_{n=0}^{\infty}(a_n\cos nx + b_n\sin nx)$, one obtains that the total power $P(z)=\int_{-\infty}^\infty \bPsi^\dagger\bPsi\, dx$ for this linear solution evolves as $P =(2\pi)^{1/2}\sum_{n=0}^\infty a_n e^{-n^2(1+4z^2)/8 }$. Thus, \rev{although $P(z)$ remains finite,} it  approaches the constant value  $\lim_{z\to\infty}P(z) = (2\pi)^{-1/2}\int_{0}^{2\pi} e^{2\alpha \eta} dx$ \rev{{\em faster than exponentially}} that is in sharp contrast with power oscillations occurring in usual periodic $\PT$-symmetric potentials~\cite{Makris11}.
 
\begin{figure}%[t]
\begin{center}
\includegraphics[width=\columnwidth]{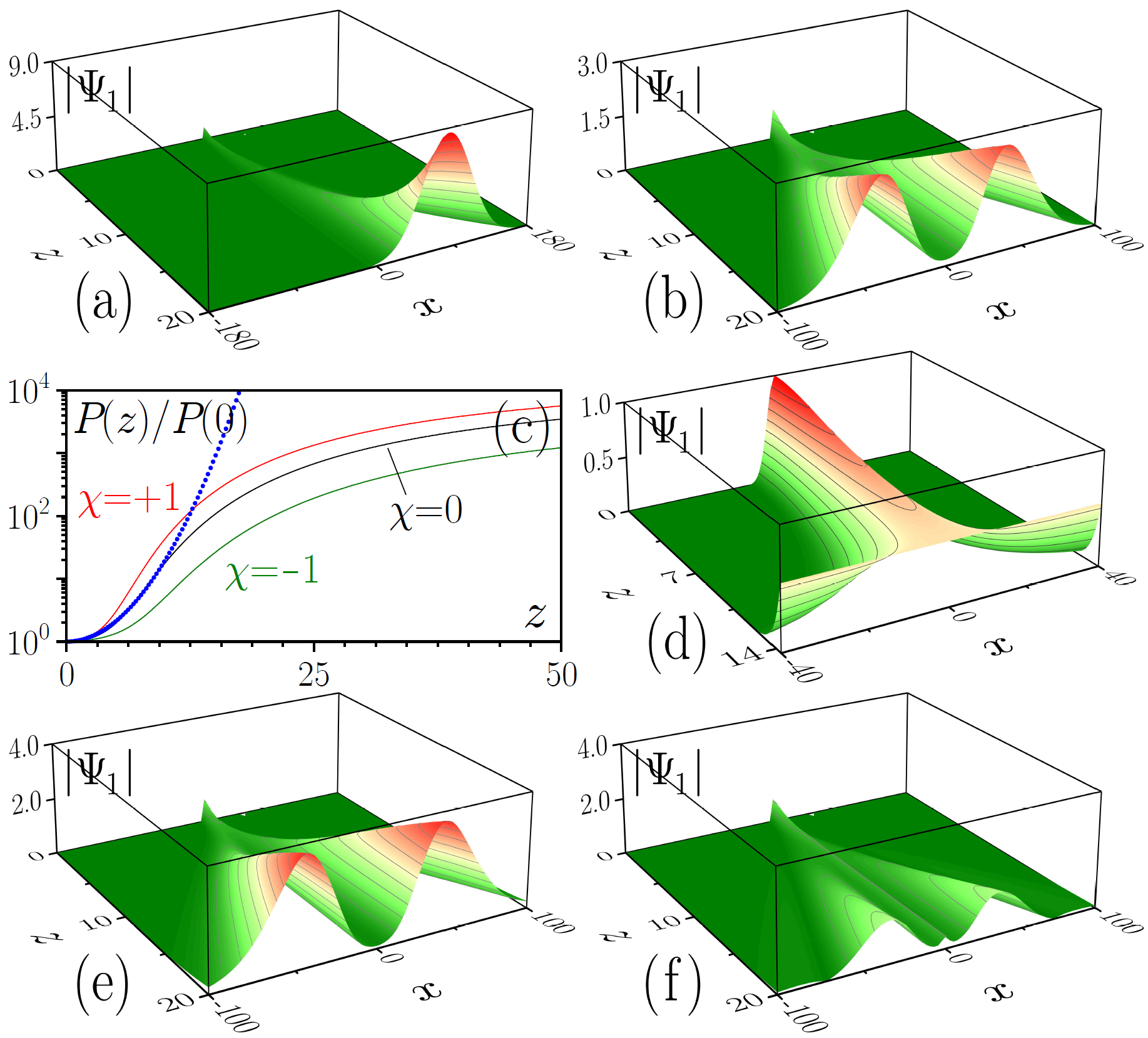}%
\end{center}
\caption{(a) Convective instability in the constant gauge field $\rA(x)=0.1 i\sigma_2$. Evolution of the beam in the gauge potential $\rA(x)=0.02ix\sigma_2/(1+0.002x^2)^2$ in the linear  (b), defocusing (e), and focusing  (f) media.  Corresponding dependencies of the power on  propagation distance are shown in (c), where labels $\chi=0, +1, -1$ correspond to linear, defocusing and focusing cases, respectively.  Blue dotted line shows  a reference   superexponential law   $\exp(0.03z^2)$. 
(d) Power blowup in the linear gauge field $\rA(x)=0.02 ix\sigma_2$ for the beam with initial width $\ell_0 \approx 2.24$. 
\rev{Panels (a,b,d) correspond to linear propagation.  Nonlinear propagation in (c,e,f) is simulated using (\ref{GPE-reduced})  where $\chi_0(x)$ is   $+1$ (defocusing medium) or $-1$ (focusing medium)  for $|x|<2$ and is zero otherwise}. Hereafter all quantities are plotted in dimensionless units}
\label{fig:one}
\end{figure}

When gauge potential (\ref{A-minimal}) is $x$-independent, i.e.,  $\eta(x)=\epsilon x$, where $\epsilon\ll 1$ guarantees the   smallness of $\eta_x$,  the total power $P(z)= P(0)e^{2(\alpha\epsilon z)^2}$ manifests {\em superexponential} growth accompanied by the directional drift of the wavepacket [Fig.~\ref{fig:one}(a)] that is a feature of {\em convective} instability~\cite{LL}. Now the system is $\p_x\T$-symmetric and its dynamics strongly contrasts with previously known giant, but bounded and periodically oscillating, amplification in a $\PT$-symmetric parabolic potential~\cite{KSZ}.

Explosive beam amplification can also be observed in spatially localized non-Hermitian gauge fields. In Fig.~\ref{fig:one}(b) we illustrate typical evolution of the beam governed by Eq.~(\ref{init}), which splits into two beams in the gauge potential $\rA(x)=0.02 i x\sigma_2/(1+\mu^2x^2)^2$, corresponding to $\alpha\eta(x)=0.01x^2/(1+\mu^2x^2)$, with $\mu\ll 1$.  \rev{For the validity of the model at large $x$ and long propagation distances $z$ in this case one has to use the auxiliary potential $U(x,y)$ in order to make $\mathfrak{U}$ in (\ref{GPE}) negligible~\cite{supplemental}, since  $|\eta_x|$ is effectively small only when the wavepacket is concentrated near $x=0$ at the initial stages of evolution.}  The power of the beam manifests quick initial growth [black curves in Figs.~\ref{fig:one}(c) and (d)] that is superexponential [in Fig.~\ref{fig:one}(d) it is approximated by the $\exp(0.03z^2)$ law shown by the blue dots]. However, this amplification is transient due to its convective character and in our example it takes place for $z\lesssim 10$.  At larger distances the beam leaves the region of localization of the gauge field.
 
\paragraph{Power blowup.} Superexponential amplification can formally be made as strong as necessary. For example, let us consider $\rA(x)=i\epsilon x\sigma_2$ [corresponding to $\alpha\eta(x)= \epsilon  x^2/2$], where $\epsilon<2$. In this case using (\ref{init}) one obtains
  \begin{eqnarray}
 \label{eq:acPsiPsi}
 \bPsi^\dagger\bPsi
 = 2(1+ 4z^2)^{-1/2} e^{-
 %\kappa(z)
 x^2/\ell^2(z)},
 \end{eqnarray}
where $\ell(z)$ characterizes the width of the beam $\ell^2(z)=  (1+ 4z^2)/[2 - \epsilon (1+ 4  z^2)]$. Solution (\ref{eq:acPsiPsi}) describes a Gaussian-shaped beam of the input width   $\ell(0)=\ell_0=(2-\epsilon)^{-1/2}$.
The specific feature of this solution is that at the {\em finite} distance 
$z=\zpb= \epsilon^{-{1}/{2}}\ell_0/(1+\epsilon\ell_0^2)$ it acquires {\em infinite width}, $\lim_{z\to\zpb}\ell(z)=\infty$, while its intensity becomes $x$-independent $\bPsi^\dagger\bPsi\to 2[1+(\epsilon\ell_0^2)^{-1}]^{-1/2}$ leading to divergence of the   power $P(z)= [1-(z/\zpb)^2]^{-1/2}P(0)$. This phenomenon is illustrated in Fig.~\ref{fig:one}(d) and it can be termed as \emph{power blowup}. 
The power blowup is characterized by the transformation of an input Gaussian beam into a constant-amplitude chirped wave. By applying time inversion $\T$ [that implies replacement of $\rA(x)$ by $\rA^*(x)$], one can show that under the action of the non-Hermitian gauge field the input chirped plane wave can be transformed into the output Gaussian beam.
 
 \paragraph{Minimal \rev{nonlinear} model.} \rev{Now  we elucidate    the effect of nonlinear terms in (\ref{GPE}).}  Due to opposite parities of the functions $\tPhi_{1}(y)$ and $\tPhi_{2}(y)$, we have
\begin{eqnarray*}
	\label{eq:tchi}
	 \tchi_1=\tchi_2^*=\chi_0\rev{(x)}:=\frac{\chi\rev{(x)}}{4}\!\!\int_{-\infty}^\infty [\tPhi_{2}(y+i\eta)-\tPhi_{1}(y+i\eta)]^3 \nonumber\\
	\times [\tPhi_{2}(y-i\eta)-\tPhi_{1}(y-i\eta)]dy.
\end{eqnarray*}
\rev{If $\eta(x)$ remains small on the support of $\chi(x)$, we can approximate}
\begin{align*}
 \chi_0\rev{(x)} \approx \frac{\chi\rev{(x)}}{4} \int_{-\infty}^\infty[\tPhi_{1}^4(y) + 6\tPhi_{1}^2(y)\tPhi_{2}^2(y) +\tPhi_{2}^4(y)]dy,
 %+\mathcal{O}(\eta^2) 
\end{align*}
where the neglected terms are proportional to $\rev{\chi(x)}\eta^2$.
Then Eq.~(\ref{GPE}) reduces to the ``minimal'' model \rev{with real effective nonlinearity}
\begin{equation}
\label{GPE-reduced}
i\frac{\partial \bPsi}{\partial z}=\frac{1}{2}\Pi^2  \bPsi +   \chi_0\rev{(x)} \left(\begin{array}{cc}
 |\Psi_1|^2 & 0 \\ 0 & |\Psi_2|^2
\end{array}\right)\bPsi.
\end{equation} 
\rev{If  $ \chi_0{(x)}$ is an even function,} then Eq.~(\ref{GPE-reduced})   with the gauge field  (\ref{A-minimal}) is $\p_x\T-$symmetric if $\eta(x)=-\eta(-x)$, and obeys $\p_x$ and  $\sigma_3\T$ symmetries if $\eta(x)=\eta(-x)$. We also note that removing the  gauge field  from   (\ref{GPE-reduced}) results in {a nonlinearity of  complex form \cite{supplemental}}.

\paragraph{Nonlinear diffraction and solitons.} The   phenomenon of power blowup resembles the well-known wave collapse  
in nonlinear media~\cite{collapse1,collapse1.5,collapse2}, in the sense that a physically meaningful solution ceases to exist at a finite blowup distance. Except for this, two phenomena are drastically different. While the usual collapse is associated with spatial contraction of the beam accompanied by the infinite growth of its amplitude for conserved total power (or $L^2$ norm), the power blowup implies the divergence of $L^2$ norm, whereas the amplitude of solution remains bounded for all $z\leq \zpb$. Hence, power blowup is a genuinely non-Hermitian phenomenon. Even more importantly, power blowup can occur in {\em linear} and effectively one-dimensional system. This raises a question about the impact of nonlinearity on the phenomenon. To address this issue we return to the   example of spatially localized gauge potential and compare the behavior of the exact linear solution with numerically simulated nonlinear propagation for the same input beam [see Fig.~\ref{fig:one}(c)]. \rev{We consider  $\chi_0(x)$ in the form of a finitely supported rectangular function, such that $\eta(x)$ remains   small within the support.} The defocusing nonlinearity, leading to faster broadening of the beam [c.f. panels (e) and (b)], results in acceleration of the initial power growth [see the red line in Fig.~\ref{fig:one}(c)]. At longer distances, however, due to convective nature of the instability, the split wavepackets propagate outwards the region of the gauge field localization [Fig.~\ref{fig:one}(e)], and the amplification in the nonlinear medium gradually slows down. For the focusing nonlinearity, at initial distances we observe slower growth of the power [Fig.~\ref{fig:one}(c) and (f)].  

\begin{figure}%[t]
\begin{center}
\includegraphics[width=0.99\columnwidth]{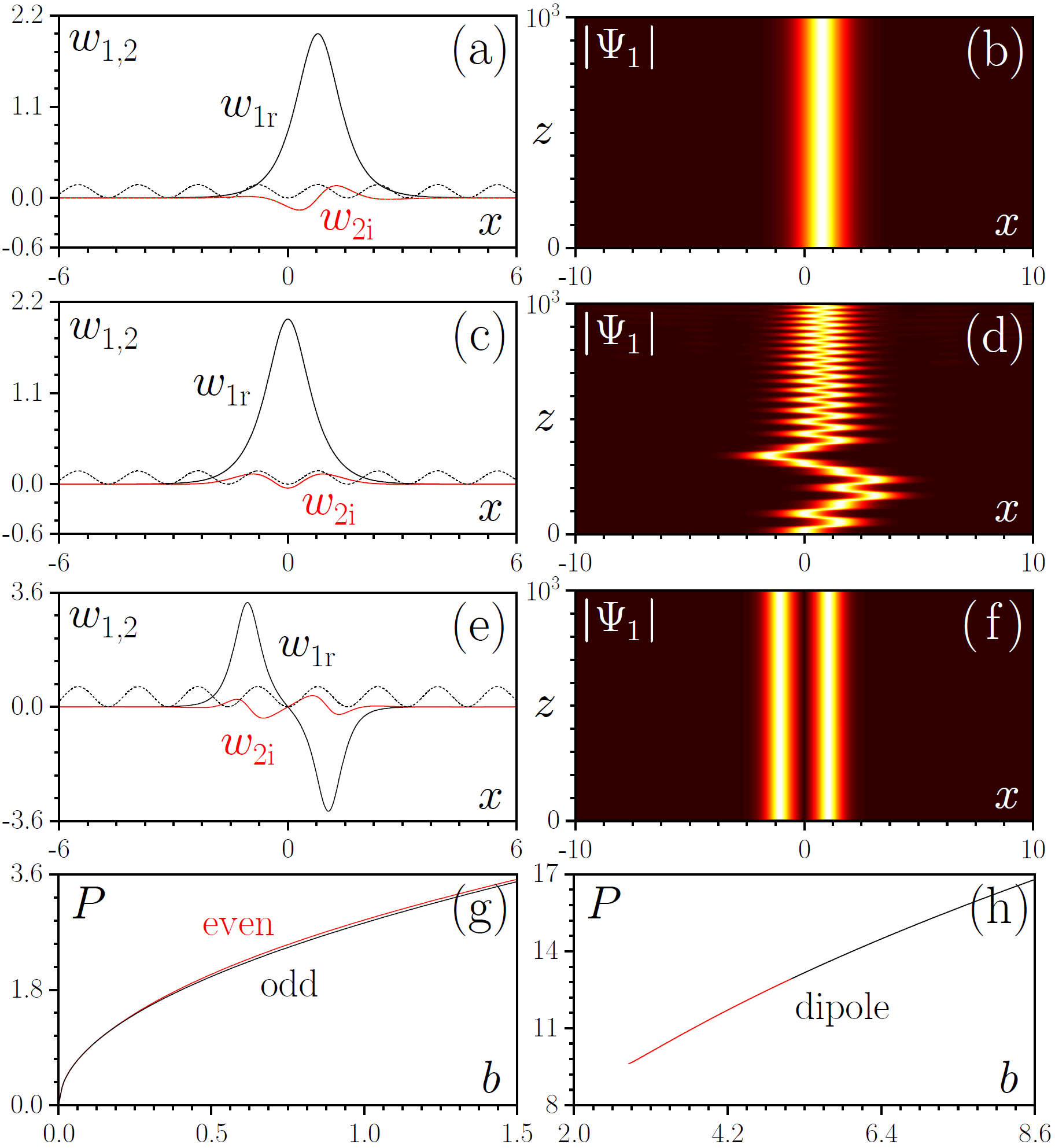}	
\end{center}
\caption{Profiles of (a) odd and (c) even solitons at  $b =2$ and $\max[\alpha\eta(x)]= 0.2$, and dipole (e) soliton at $b =5.5$ and $\max[\alpha \eta(x)]= 0.4$. Only nonzero components $w_{1r}$, $w_{2i}$  are shown. The dashed line shows $\alpha^2 \eta_x^2$. Propagation of these solitons is shown in (b), (d), and (f), respectively. (g) Power of odd and even solitons {\it versus} $b$ at $\max[\alpha\eta(x)]=0.2$. (h) Power of dipole soliton {\it versus} $b$ at $\max[\alpha\eta(x)]= 0.4$. Stable (unstable) branches are shown black (red). \rev{All results correspond to the uniform focusing nonlinearity $\chi_0(x)=-1$.}}
	\label{fig:soli}
\end{figure}

Although our system is non-Hermitian, 
the presence of the $\p_x$ and $\sigma_3\T$ symmetries, discussed above,   suggests that the nonlinearity can enable   families of bright solitons~\cite{KYZ,Suchkov2016}. In Fig.~\ref{fig:soli} we illustrate such families for a representative example of periodic gauge field $\eta(x) = \eta_0\cos(2x)$. Solitons of Eq. (\ref{GPE-reduced}) can be found in the form $\Psi_{1,2}=e^{ibz}w_{1,2}(x)$, with real propagation constant $b$ and $w_k=w_{k{\rm r}}+iw_{k{\rm i}}$. Importantly, inhomogeneous gauge potential dictates stable equilibrium positions for soliton center. Thus, fundamental bell-shaped solitons can be stable only if they reside on maxima of the $\eta_x^2$ function [see Figs.~\ref{fig:soli}(a,b) corresponding to such ``odd'' states], while solitons residing on minima of $\eta_x^2$ exhibit drift instabilities [Figs.~\ref{fig:soli}(c,d), ``even'' states]. Moreover, non-Hermitian gauge potentials arrest repulsive forces between out-of-phase solitons leading to formation of dipole and more complex solitons [Figs.~\ref{fig:soli} (g,h)]. Dipole solitons can also be stable (at least for certain intervals of the propagation constant), if the amplitude of the  gauge field is large enough.

\begin{figure}%[t]
\begin{center}
\includegraphics[width=1.00\columnwidth]{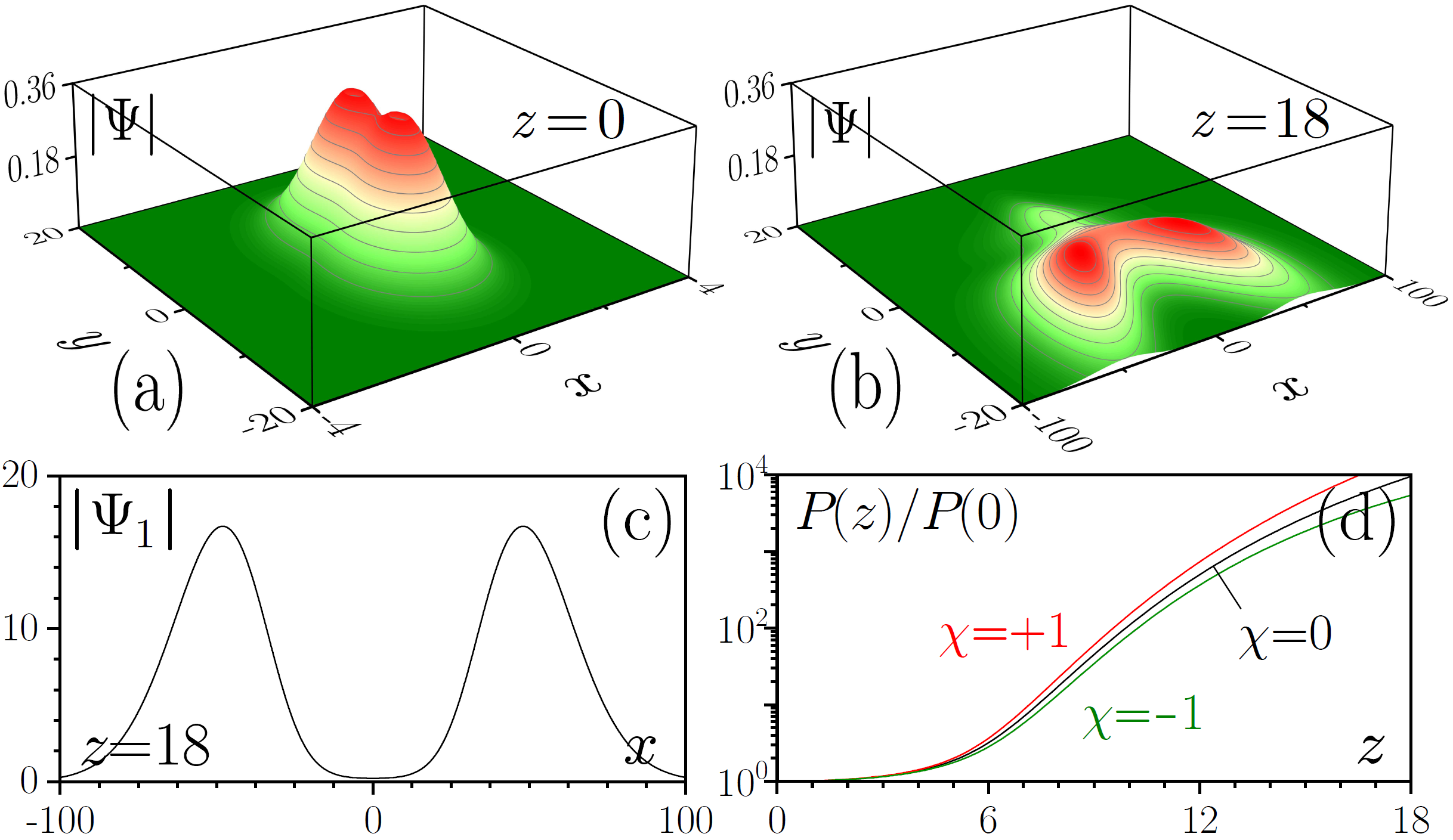}	
\end{center}
\caption{\rerev{2D field distribution at $z =0$ (a) and $z =18$ (b). The field at $z =0$ corresponds to the input of 1D simulation in Fig. 2(b). (c) Amplitude of the envelope $\Psi_1$ extracted from the 2D field distribution at $z =18$. (d) Power of
the spinor envelope $(\Psi_1 , \Psi_2)^T$  extracted from 2D simulations for zero, defocusing and focusing nonlinearities. Note that, in spite of the huge power amplification in (d), the amplitude and power of the (2+1)D field $\Psi$ in (a,b) do not grow.}}
	\label{fig:2D}
\end{figure}

\rerev{\paragraph{Evolution of the 2D field.} To validate the approximations used for derivation of the reduced (1+1)D model (\ref{GPE}), we have studied evolution in the original full (2+1)D equation  (\ref{opt-eq1}) for the potential $V\equiv V_{\rm ex}(y+i\eta(x))$, defined above (see also \cite{razavy1980,supplemental}), without the additional potential $U(x,y)$. Figure~\ref{fig:2D} presents the 2D results for the same gauge potential as the one used in Fig.~\ref{fig:one}(b,c,e,f). The input 2D field $\Psi(x,y,z=0)$ [Fig.~\ref{fig:2D}(a)] is constructed from the initial conditions used in Fig.~\ref{fig:one}(b,c,e,f).  Propagation of the 2D field is simulated using Eq. (\ref{opt-eq1}), and then the 1D spinor wavefunction is extracted from the 2D data using the projection $\Psi_{1,2}(x,z)=\int_{-\infty}^\infty \phi_{1,2}(x,y) \Psi(x,y,z) dy$.  
The results for the spinor field  shown in Fig.~\ref{fig:2D}(c,d) are in good qualitative agreement with the predictions obtained from the reduced model. In particular, the full simulation reproduces the beam  splitting [compare Fig.~\ref{fig:one}(b,e,f) and Fig.~\ref{fig:2D}(b)] and transient superexponential amplification for the power  $P(z)$ of the (1+1)D spinor,  in linear and nonlinear regimes [compare Fig.~\ref{fig:one}(c) and Fig.~\ref{fig:2D}(d)]. Remarkably, in spite of the huge growth of $P(z)$, the amplitude of the field (2+1)D field $\Psi$ itself does \emph{not} undergo the superexponential growth. In other words, the superexponential growth is \emph{emulated} by the effective (1+1)D model due to the non-normalized eigenstates $\phi_{1,2}$, whereas the real optical power does not grow appreciably (i.e., the phenomenon is indeed experimentally feasible).}

To conclude, we introduced a system of two optical waveguides emulating non-Hermitian matrix gauge potential. Field propagation in this setup  features unusual properties even in the linear regime. These are superexponential amplification and finite-distance power blowup, accompanied by   complete spatial delocalization of the wavepacket. While in the linear model the non-Hermitian matrix field can be gauged out, in Kerr media this transformation leads to an inhomogeneous non-Hermitian nonlinearity which supports families of fundamental and dipole vector solitons. The approach  can be directly generalized to multiple-waveguide optical systems and to gases of multilevel atoms, thus allowing design of non-Abelian non-Hermitian gauge fields in higher-dimensional settings.

\begin{acknowledgments}
VVK is grateful to I. V. Barashenkov for useful comments. The work of DAZ was supported by the Foundation for the Advancement of Theoretical Physics and Mathematics ``BASIS'' (Grant No. 19-1-3-41-1). VVK acknowledges financial support from the Portuguese Foundation for Science and Technology (FCT) under Contract no. UIDB/00618/2020.
\end{acknowledgments}

\end{document}